# Strain gradient induced polarization in graphene

S. I. Kundalwal, [a, #] S. A. Meguid, [a,]* and G. J. Weng [b]

[a]Mechanics and Aerospace Design Laboratory, Department of Mechanical and Industrial Engineering, University of Toronto, Toronto, Ontario M5S 3G8, Canada

[b]Department of Mechanical and Aerospace Engineering, Rutgers University, New Brunswick, NJ 08903, USA

## Abstract

Flexoelectricity phenomenon is the response of electric polarization to an applied strain gradient and is developed as a consequence of crystal symmetry in all materials. In this study, we show that the presence of strain gradient in non-piezoelectric graphene sheet does not only affect the ionic positions, but also the asymmetric redistribution of the electron density, which induce strong polarization in the graphene sheet. Using quantum mechanics calculations, the resulting axial and normal piezoelectric coefficients of the graphene sheet were determined using two loading conditions: (i) a graphene sheet containing non-centrosymmetric pores subjected to an axial load, and (ii) a pristine graphene sheet subjected to a bending moment. Particular emphases were placed on the role of edge and corner states of pores arising due to the functionalization. We also investigated the electronic structure of graphene sheet under different in-plane strain distributions using quantum mechanics calculations and tight-binding approach. The findings of our work reveal that the respective axial and normal electromechanical couplings in graphene can be engineered by changing the size of non-centrosymmetric pores and radii of curvature. Our fundamental study highlights the possibility of using graphene sheets in nanoelectromechanical systems as sensors or actuators.

[#]Currently Assistant Professor in the Department of Mechanical Engineering, Indian Institute of Technology Indore, Indore 453 552, India. E-mail: kundalwal@iiti.ac.in.
*Corresponding author. Tel.: +1 (416) 978 5741; Fax: +1 (416) 978 7753. E-mail: meguid@mie.utoronto.ca (S.A. Meguid).

# 1. Introduction

Recent advances in nanoscale technologies have renewed interest in flexoelectricity. Large strain gradients present at the nanoscale level may lead to strong electromechanical coupling. Piezoelectricity - electrical polarization induced by a uniform strain (or vice-versa) - is the most widely known and exploited forms of electromechanical coupling that exists in non-centrosymmetric crystals. In centrosymmetric crystals, the presence of the center of inversion results in the absence of bulk piezoelectric properties. In contrast to piezoelectricity, flexoelectricity can be found in any crystalline material, regardless of the atomic bonding configuration [1]. The symmetry breaking at surfaces and interfaces in nonpolar materials would allow new forms of electromechanical coupling, such as surface piezoelectricity and flexoelectricity, which cannot be induced in bulk materials. The first step towards a theoretical understanding of flexoelectricity was due to Kogan [2]. It is worth mentioning that the term "flexoelectricity" for crystalline materials was coined by a similar phenomenon in liquid crystals [3]. Piezoelectric materials are commonly used where precise and repeatable controlled motion is required, such as atomic force microscopy probes, smart structures [4-6], piezotronic devices [7], sensors and actuators, and nanogenerators [8]. The conventional piezoelectric ceramics are heavy, brittle, and some pose significant environmental risk due to the inclusion of high lead content [9]. Piezoelectric polymers, on the other hand, are light and environmentally benign, but typically display considerably smaller piezoelectric response in terms of actuation. Piezoelectric single-layer hexagonal boron nitride is a promising nanomaterial for nanoelectromechanical system (NEMS), but its synthesis is much harder which has been a major issue for more than a decade.

In view of their extraordinary mechanical, thermal, and physical properties, CNTs have emerged as one of the most promising nano-reinforcements that can also be used to tailor the properties of polymers [10-20]. Moreover, the successful synthesis of 2D single layer graphene sheet [21] has attracted considerable attention from both academia and industry. This is due to its unique scale-dependent electronic, mechanical, and thermal properties [22-25]. It is widely considered to be one of the most attractive materials of the twenty-first century. Remarkably, high electron mobility and other striking features, such as anomalous quantum hall effect, pseudomagnetic field, and spin transport [26], have made graphene an intensely studied material from both basic science viewpoint as well as advanced applications in micro- and nano-electronics [7], spintronic devices, electron lenses, energy storage systems [27], and gas separation membranes [28]. In view of its unique 2D structure and electrical properties with zero bandgap, graphene is most suitable for NEMS applications.

The flexoelectric effect in carbon nanotubes was observed by White et al. [29] for the first time in 1993. In their study, the bond symmetry breaking due to curvature was visible in the electronic properties of carbon nanotubes (CNTs). A homogeneous mechanical deformation of graphene cannot induce polarization due to the symmetry of its lattice. However, second-order electronic flexoelectric effect in graphene can be induced by the presence of a strain gradient. The strain gradient changes the ionic positions as well as leads to asymmetric redistribution of the electron density. Contrary to 3D systems, graphene are also able to sustain large strains up to 25% [24] and, thus can exhibit exceptional forms of electromechanical coupling [30]. Zeinalipour-Yazdiand and Christofides [31] determined the size-dependent Young's modulus and C-C binding energy in graphene nanoribbons via electron first principles computations. They reported linearity between the C-C binding energy and Young's modulus, which was explained by the decrease in the molecular polarizability due to deformations within the proportional limit of the graphene nanoribbons. Dumitrică et al. [32] investigated the normal polarization induced by bending of graphite shells, which microscopically occurs because of a shift in $sp^2$ hybridization at each atomic site. Using density functional theory (DFT) calculations for bent nanographitic ribbons made of up to 400 atoms, Kalinin and Meunier [33] predicted the electromechanical coupling, in agreement with the prediction of linear theory for flexoelectric coupling in quantum systems. Note that the stability of planar carbon materials is higher than their curved nanostructures (e.g. Fullerenes and CNTs) and this is associated to the existence of the energy requirements to bend a graphene sheet. Based on average bond-dissociation-enthalpies, Zeinalipour-Yazdi and Loizidou [34] applied physical sphere-in-contact models to elucidate the cap geometry of single walled CNT with sub-nanometer width (i.e., (3,3), (4,4) and (5,5)-SWCNTs). Their DFT calculations reveal that these new carbon geometries have similar stability to carbon found in other ball-capped zigzag and arm-chair nanotubes (i.e., (5,5), (6,6), (9,0) and (10,0)). Recently, experimental and theoretical investigations reported that creating nanoscale non-centrosymmetric pores in graphene nanoribbons would lead to apparent piezoelectric behavior [35, 36]. These exceptional and highly desirable electromechanical properties have motivated our interest in this work. To the best of the authors' knowledge, the axial and normal flexoelectricity coefficients of graphene have not yet been reported. In this study, we first examined the band structures of graphene sheet subjected to different in-plane strain distributions using quantum mechanics calculations and tight-binding (TB) approach. Then, we focused our attention to determining the induced piezoelectric coefficients of graphene with and without non-centrosymmetric pores of different sizes using quantum mechanics calculations. The outcome of this work should lead to a greater insight into strain-induced electric polarization in graphene sheets that would allow exploring the essence of their strong piezoelectric response through different loadings.

## 2. TB modeling of electronic structure of graphene

Graphene has unique electronic properties evolving from its hexagonal honeycomb lattice structure, which makes electrons in graphene behave as massless relativistic fermions that satisfy the Dirac equation [37]. In-plane strain distributions in graphene sheet significantly modify its band structure around the Fermi level, which break the inversion symmetry [38]. Theoretically, both the TB model and ab initio approaches have been widely adopted to investigate the effect of strain on the band structure of graphene and graphene nanoribbons [39, 40]. In this research, we first study the electronic structure of graphene sheet under different in-plane strain distributions using quantum mechanics calculations and Hückel TB model [40, 41]. If the inversion symmetry breaking is confirmed with particular planer strain distribution case, then the flexoelectric effect in graphene sheet with non-centrosymmetric pore is examined.

We define x-direction to be along the unstrained graphene axis and y as the transverse direction, as shown in Fig 1.

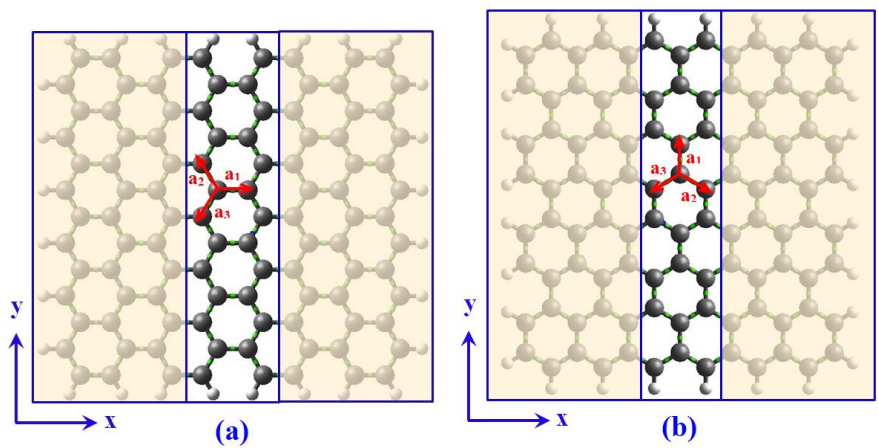

**Fig. 1.** The unit cells of **(a)** armchair graphene sheet, and **(b)** zigzag graphene sheet. $a_1$, $a_2$ and $a_3$ are the bond vectors.

The following changes occur in the x-component of the coordinate of the $i^{th}$ atom when the graphene sheet is subjected to uniaxial or shear strains:

$$\left.\begin{array}{l} x_i \to (1+\varepsilon_A)x_i \\ x_i \to (1+\varepsilon_z)x_i \end{array}\right\} \quad \text{Uniaxial strains} \quad (1a)$$

$$x_i \to x_i + \gamma y_i \quad \text{Shear strain} \quad (1b)$$

where $\varepsilon_A$ and $\varepsilon_z$ are the respective uniaxial strains along armchair and zigzag directions, and $\gamma$ is the shear strain. The uniform strain in graphene sheet can be written as a strain tensor:

$$\boldsymbol{\varepsilon} = \begin{bmatrix} \varepsilon_A & \gamma \\ 0 & \varepsilon_z \end{bmatrix} \quad (2)$$

In the deformed graphene sheet, real space vectors are

$$\mathbf{r} = (\mathbf{I} + \boldsymbol{\varepsilon})\mathbf{r}_0 \quad (3)$$

in which $\mathbf{I} = (\delta_{ij})_{2\times 2}$ and subscript "0" denotes the undeformed state of graphene sheet.

The resulting Brillouin zone (BZ) for the reciprocal vector $\mathbf{k}$ in the deformed space is no longer a regular hexagon [39]. By introducing a new quantity

$$\mathbf{k}' = (\mathbf{I} + \boldsymbol{\varepsilon})^T \mathbf{k} \quad (4)$$

The quantity $\mathbf{k}'$ can be regarded as the undeformed reciprocal vector. The BZ in the $\mathbf{k}'$ space is restored to hexagonal because

$$\mathbf{k}\mathbf{r} = \mathbf{k}(\mathbf{I} + \boldsymbol{\varepsilon})\mathbf{r}_0 = (\mathbf{I} + \boldsymbol{\varepsilon})^T \mathbf{k}\mathbf{r}_0 = \mathbf{k}'\mathbf{r}_0 \quad (5)$$

This approach facilitates analysis of variations in electronic states near Fermi point $\mathbf{k}_F$ with strains and the Hückel TB Hamiltonian becomes

$$H(\mathbf{k}) = \sum_{i=1,2,3} t_i \exp(i\mathbf{k}\mathbf{a}_i) = \sum_{i=1,2,3} t_i \exp(i\mathbf{k}'\mathbf{a}_{i0}) \quad (6)$$

with Harrison hopping parameter $t_i = t_0(\mathbf{a}_0/\mathbf{a}_i)^2$ [42]. This parameter incorporates the effect of strain into the variation of $t_i$. Under the application of symmetric strains i.e., $t_1 = t_2 = t_3$ (see

Fig. 2a), the Fermi point $\mathbf{k}_F$ can be determined by solving $E(\mathbf{k}_F) = |H(\mathbf{k}_F)| = 0$. On the other hand, $t_1$, $t_2$, and $t_3$ are asymmetric in case of uniaxial strains, as shown Figs. 2(b) and 2(c), and the Fermi points deviate from K in the $\mathbf{k}'$ space.

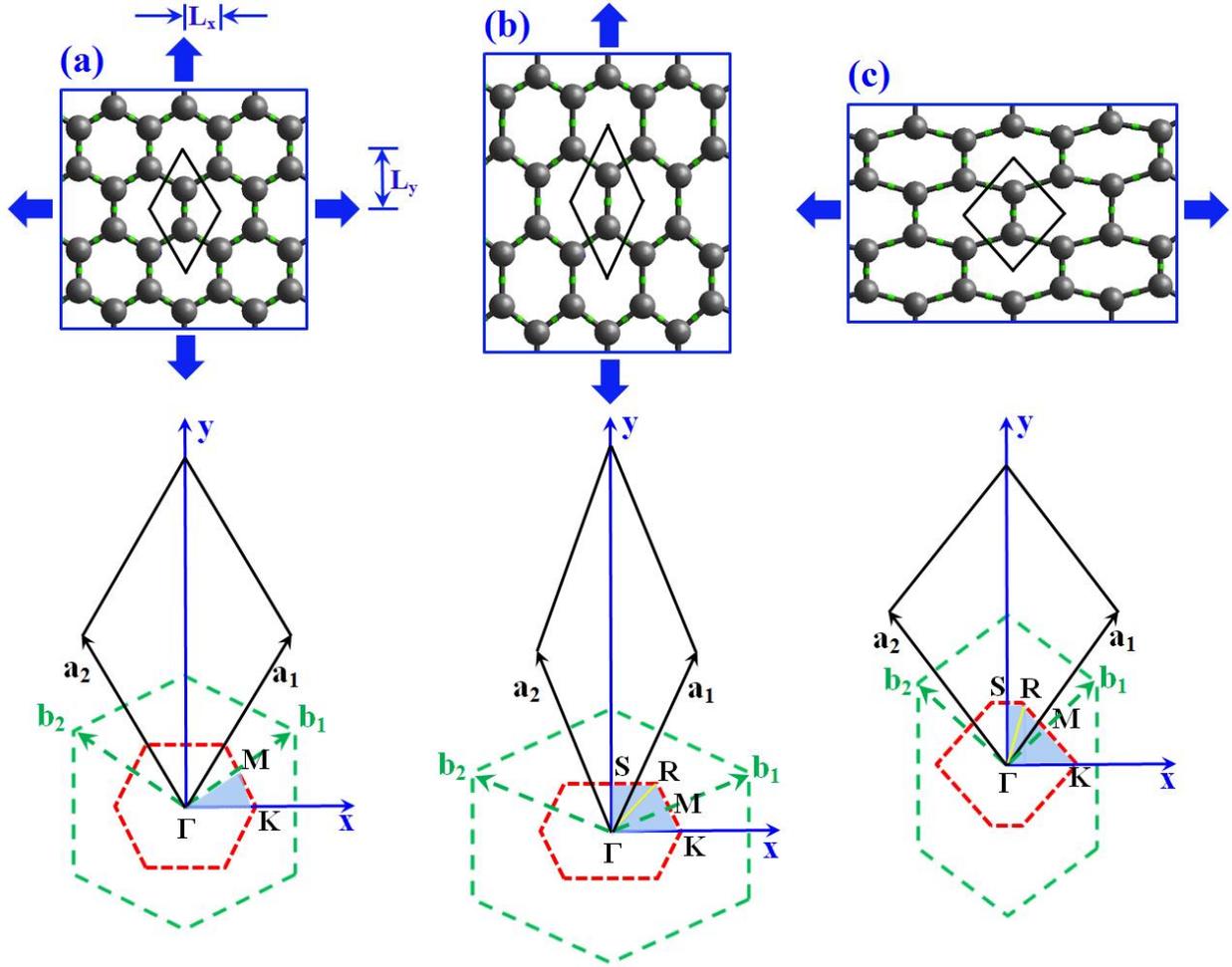

**Fig. 2.** **(a)** Graphene system with symmetrical strain distribution, **(b)** armchair graphene with asymmetrical strain distribution, and **(c)** zigzag graphene with asymmetrical strain distribution. Corresponding primitive cells in black solid line, reciprocal lattices in green dashed line, BZs in red dashed line, and irreducible BZs in sky blue color are shown below the deformed graphene systems. Γ, K, M, R, and S are the high symmetrical points; and $L_x$ and $L_y$ are the half diagonal lengths of the primitive cells.

Under the linear approximation, the deviation $\Delta \mathbf{k}'_F = \mathbf{k}'_F - \mathbf{k}'_K$ is determined to be [41]:

$$\Delta \mathbf{k}'_{Fx} a_0 = C_t [\varepsilon_y [(1 + \upsilon)\cos 3\theta + \gamma \sin 3\theta]] \tag{7a}$$

$$\Delta \mathbf{k}'_{Fy} a_0 = -C_t [\varepsilon_y (1 + \upsilon)\sin 3\theta + \gamma \cos 3\theta] \tag{7b}$$

where $\Delta \mathbf{k}'_{Fx}$ and $\Delta \mathbf{k}'_{Fy}$ are the respective components of $\Delta \mathbf{k}'_F$; $\theta$ is the angle between the x- axis and the zigzag direction, $\varepsilon_y$ is the uniaxial strain along the y- direction, $\upsilon$ is Poisson's ratio, $a_0$ is the equivalent bond length and $C_t$ is given by

$$C_t = -\frac{a}{2t}\frac{dt}{da}\bigg|_{a=a_0} \tag{8}$$

The deviation of K′ is opposite to that of K in the $\mathbf{k}'$ space. The dispersion relation of the deformed graphene is given by expanding E($\mathbf{k}$) near the Fermi points:

$$E(\mathbf{k}') = \pm \frac{3}{2} t_0 a_0 |\mathbf{k}' - \mathbf{k}'_F| \tag{9}$$

Thus, the effect of small strain is to move the energy cone in the $\mathbf{k}'$ space, as demonstrated in Fig. 3.

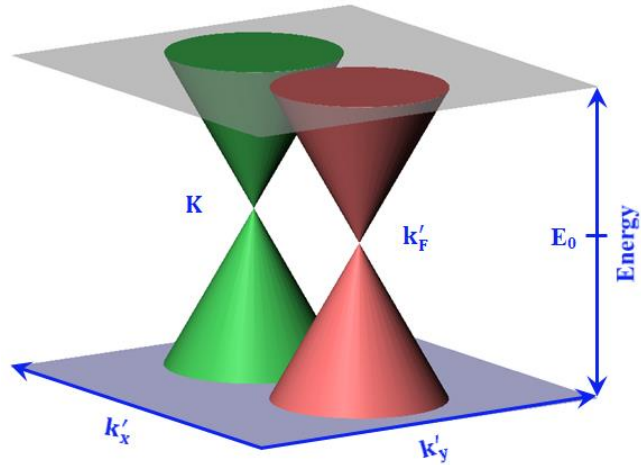

**Fig. 3.** Schematic illustration of shift of energy cone away from K points in $\mathbf{k}'$ space.

Combining Eqs. (7) and (9), we can obtain the following relations for the pseudo gap at K and K′ points in the **k**′ space:

$$E_{gap}(k'_V) = 3t_0 S_t (1 + \upsilon_A)\varepsilon_A \tag{10a}$$

$$E_{gap}(k'_V) = 3t_0 S_t (1 + \upsilon_z)\varepsilon_z \tag{10b}$$

## 3. Quantum mechanics calculations

### 3.1. Density Functional Theory

DFT is an attractive method when compared with conventional wave function based methods, because the electronic wave function of a system with N-electrons depends on 3N spatial coordinates, while the electron density depends only on three coordinates (x, y and z). In DFT, the total electronic energy $E(\rho)$ is the sum of the kinetic energy of the electrons $T(\rho)$, the energy of the electrons in the external field $V(\rho)$, and the electron-electron interaction energy $V_{e-e}(\rho)$; viz.,

$$E(\rho) = T(\rho) + V(\rho) + V_{e-e}(\rho) \tag{11}$$

In 1964, Hohenberg and Kohn proved [43] that the energy and electronic properties of atoms and molecules can be determined by the electron density $\rho(\mathbf{r}) = \rho(x, y, z)$. Using their theory, the properties of a many-electron system can be characterized using functionals, i.e. the spatially dependent electron density. The Kohn-Sham equation in Hartree atomic units [44] is given by

$$\left[-\frac{\nabla^2}{2} + \vartheta_{ext}(\mathbf{r}) + \vartheta_{Hartree}[\rho](\mathbf{r}) + \vartheta_{XC}[\rho](\mathbf{r})\right]\psi_i(\mathbf{r}) = \psi_i(\mathbf{r}) \tag{12}$$

where the first term represents the kinetic energy of electrons, the second is the external potential, and the third is the Hartree potential that describes the classical electrostatic repulsion between the electrons. The XC potential $\vartheta_{XC}[\rho]$ is defined by

$$\vartheta_{XC}[\rho](\mathbf{r}) = \frac{\delta E_{XC}[\rho]}{\delta \rho(\mathbf{r})} \tag{13}$$

where $E_{XC}[\rho]$ is the XC energy functional associated with the electronic density ρ defined by

$$\rho(\mathbf{r}) = \sum_{i}^{\text{Occupied}} |\psi_i(\mathbf{r})|^2 \qquad (14)$$

where the sum runs over the occupied states.

*3.2. Band gap and flexoelectric effect in graphene*

In this section, all electron quantum mechanical calculations based on DFT were developed using Quantum ESPRESSO software package [45] and ultrasoft pseudopotentials. We considered both symmetric and asymmetric in-plane strain distributions in graphene sheets (see Fig. 2) and calculated their band gaps. In our simulations, we keep one of $\varepsilon_A$, $\varepsilon_z$ and $\gamma$ at a fixed value, while the other two are relaxed to achieve the lowest total energy. The edges in graphene sheet were passivated using hydrogen because they both result in dangling bonds that lead to extended reconstructions [46]. First, the initial configuration of graphene sheet was prepared and then relaxed under the application of zero and finite strain conditions. Once the graphene system was fully relaxed, electronic band structure calculations were carried out to ensure the semiconducting nature of its geometry. The quantum mechanics calculations were performed in an unstrained reference state and under various levels of strain up to 3% in the directions as shown in Fig. 2. The geometry of graphene was optimized until the residual force per atom becomes less than 0.04 eV/Å and then the electronic ground state wave functions were collected using a mesh of 45 × 45 Monkhorst-Pack k-points for k-point sampling in the BZ. The energy was considered to be converged once the change in the total energy of the system between subsequent steps was less than 0.1 meV. Poisson ratios $\varepsilon_A$ and $\varepsilon_z$ in the directions parallel to or perpendicular to C-C bonds were calculated during each level of strain using their respective definitions; viz.,

$$\varepsilon_A = \frac{\Delta L_x/L_x}{\Delta L_y/L_y} \quad \text{and} \quad \varepsilon_z = \frac{\Delta L_y/L_y}{\Delta L_x/L_x} \tag{15}$$

The constitutive relation for the polarization vector induced due to the flexoelectricity effect may be written as [2]:

$$P_i = e_{ijk}\varepsilon_{jk} + f_{ijkl}\frac{\partial \varepsilon_{kl}}{\partial x_j} \tag{16}$$

where $e_{ijk}$ and $f_{ijkl}$ are the respective piezoelectric and flexoelectric tensors; $\varepsilon_{jk}$ and $\frac{\partial \varepsilon_{kl}}{\partial x_j}$ are the elastic strain and strain gradient, respectively. Note that the first term in the right-hand side of Eq. (16) is the well-known piezoelectric effect.

Graphene sheets with non-centrosymmetric pores act as a dielectric material and under the application of axial stress it shows flexoelectricity effect [35]. This is attributed to the fact that the differences in material properties at the interfaces result in the presence of strain gradients under the application of an axial stress, which eventually induces the necessary polarization. Note that certain symmetry rules for the pores should be followed to induce polarization in the graphene sheet. Curvature-induced polarization is another way to induce flexoelectric effect in the graphene. As a result of bending of graphene, Coulomb repulsion inside a cavity increases with curvature and it leads to a redistribution of the π-orbitals. This results into an electronic charge transfer from the concave to the convex region, and induces the normal atomic dipole at each atomic site [32]. We use both the loading conditions to obtain the induced axial and normal piezoelectricity coefficients through quantum mechanics calculations. In the following sections, we will investigate how these loading conditions induce flexoelectricity phenomena in graphene systems. In addition, we will also consider different pores geometries of varying sizes to investigate their effect on any induced flexoelectric effects in graphene sheets.

## 3.3. Calculation of induced axial piezoelectric coefficient

We performed DFT calculations as those described in Section 2.2 to obtain the apparent piezoelectric behavior from a graphene sheet containing non-centrosymmetric trapezoidal shaped pore (see Fig. 4).

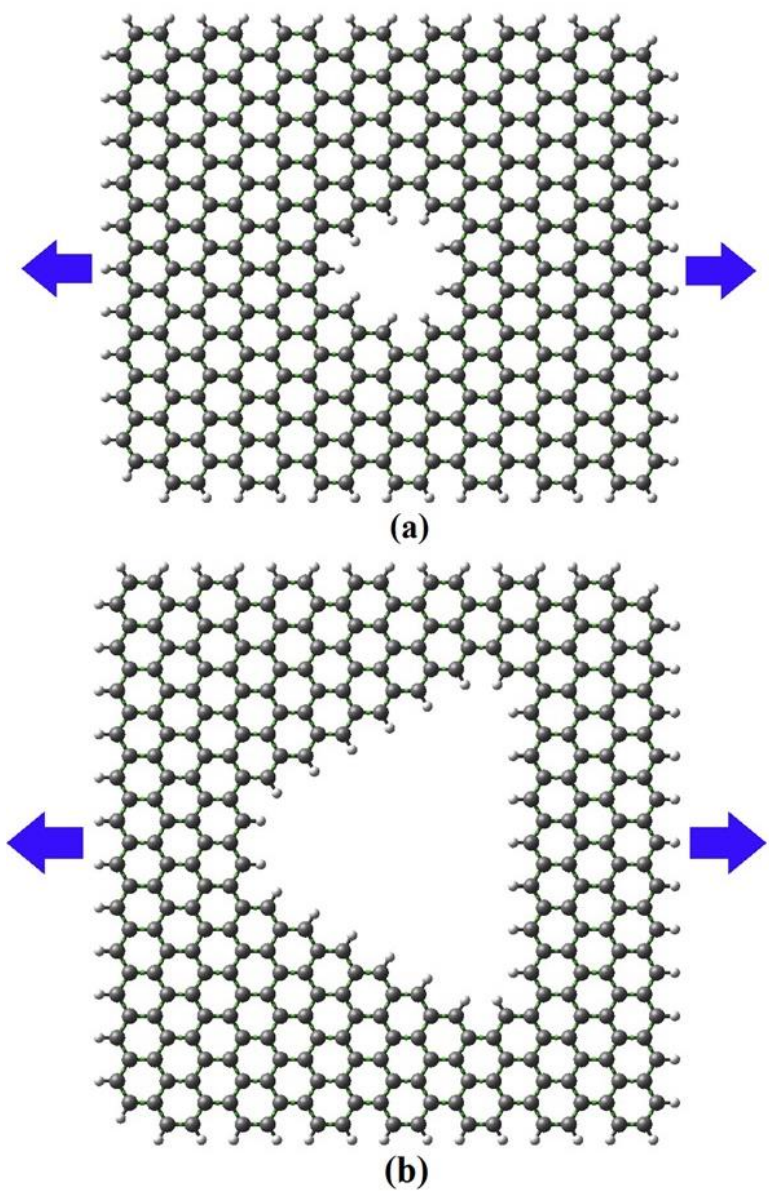

**Fig. 4.** Passivated armchair graphene sheet with trapezoidal pore subjected to an axial stress: (a) atom vacancy concentration of 4% and (b) atom vacancy concentration of 19.5%.

The edges and porosity in a graphene sheet were passivated using hydrogen and the electronic ground state wave functions were collected using a mesh of 1 × 6 × 6 Monkhorst-Pack k-points for k-point sampling in the BZ. Spacing of 0.1 Å$^{-1}$ k-point was used during the relaxation of the graphene sheet. The Berry-phase method was employed to compute the piezoelectric properties of graphene, which are directly related to the polarization differences between its strained and unstrained states. In the modern theory of polarization, the spontaneous electric polarization in periodic solids is obtained from the geometric phase (i.e., Berry phase) accumulated by the occupied electronic states as one introduces a potential that adiabatically connects an unpolarized and a polarized states of the system [47]. For graphene sheet, the Berry phase and hence the polarization is controlled by the periodic boundary conditions of the electronic wave functions. The average induced polarization was obtained for a strained graphene using the following Berry phase formulation (discretized on a dense k-point grid along the direction of the polarization):

$$\langle e_{11} \rangle = e_o + d\varepsilon + O(\varepsilon^3) \qquad (17)$$

where $e_o$ is the pre-existing polarization due to surface effects present in the unstrained state. In our simulations, we ensured the application of very small values of strain and the term ($O(\varepsilon^3)$) exhibiting nonlinear effects was neglected.

*3.4. Calculation of induced normal piezoelectric coefficient*

In this section, we determine the normal polarization induced by the bending of a graphene sheet. Note that there is no dipole moment across the flat graphene sheet due to the symmetry of π-orbitals (see Fig. 5a). By bending the graphene sheet, we can introduce asymmetry in the π-orbital overlap (see Fig. 5b).

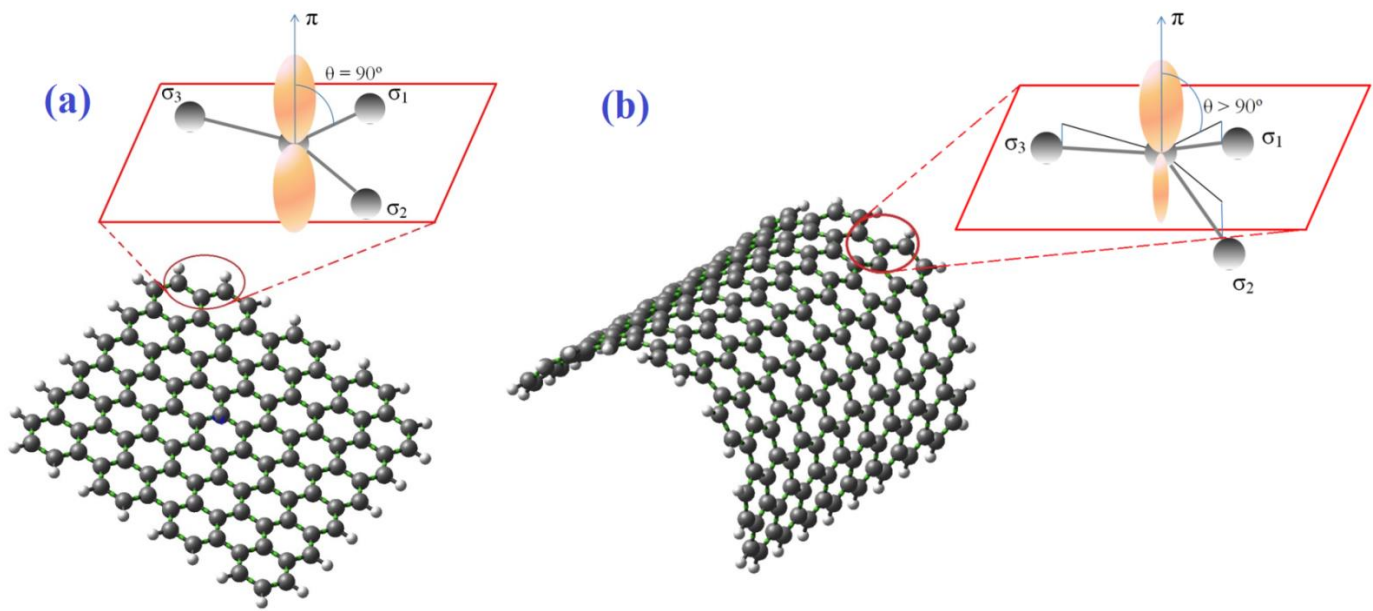

**Fig. 5.** Graphene sheets in which π-orbitals are **(a)** symmetric, and **(b)** asymmetric.

Therefore, Coulomb repulsion inside the cavity increases with curvature and leads to a redistribution (rehybridization) of the π-orbitals from the $sp^2$ of graphene to something intermediate between $sp^2$ and $sp^3$ [32]. This results into an electronic charge transfer from the concave to the convex region. Figure 5(b) demonstrates that the carbon atoms in the bent graphene sheet are not in the tangential plane and the three $\sigma_i$ – bonds (i = 1, 2, 3) are tilted down with respect to that tangential plane. In that configuration, the $\sigma_i$ – bonds are forced to lie along the internuclear axes and we can use orbital orthogonality to solve the π-orbital hybridization and direction. Due to the geometrical tilting of σ-bonds, π- and s-orbital mixing occurs which we call it π-hybrids. As shown in Fig. 5(b), the angle between the σ and π hybrids has the respective values of $109^0$ and $90^0$ for $sp^3$ and $sp^2$ hybridization. Such π- and s-orbital mixing is also reported to occur in case of conjugated polymers such as polyenes, polyacenes, and polyynes [48]. Redistribution of ions and charges occur upon bending of graphene sheet which in turn results in the formation of a net dipole moment across the graphene. Such dipole moment (μ) was obtained by fitting the energy curve E (ε) as a function of electric field (0, 0, $\varepsilon_z$) by

$$E = E_o + \mu\varepsilon_z + e_{33}\varepsilon_z^2 + O(\varepsilon_z^3) \tag{18}$$

where $E_o$ is the pre-existing energy present in a flat graphene state.

## 4. Results and discussions

We first investigated the electronic structure of a graphene sheet subjected to different in-plane strain distributions using quantum mechanics calculations and TB approach. We considered both armchair and zigzag graphene sheets. In case of zigzag graphene sheets, edge magnetism may be an issue, but we did not consider magnetic edge states in the current study because they do not exist in real passivated graphene systems. Kunstmann et al. [49] critically discussed the stability of edge states and magnetism in zigzag edge graphene nanoribbons and showed that there are at least three very natural mechanisms. They defined these three states as edge reconstruction, edge passivation, and edge closure—which dramatically reduce the effect of edge states in zigzag edge graphene nanoribbons or even totally eliminate them. Therefore, edge magnetism can be neglected in case of passivated graphene sheets. The anisotropic nature of the graphene sheet was observed by the varied Poisson's ratios under asymmetrically large strains for armchair and zigzag graphene sheets, as shown in Fig 6. This is attributed to the significant changes of the band-gap structure, the electron redistribution due to the induced large strains and the variation of angles of two adjacent C−C bonds of a graphene sheet. The angles of C−C bonds change when a graphene sheet is subjected to uniaxial strains along the armchair and the zigzag directions, which eventually influence Poisson's ratios. Our DFT results concerning the anisotropic behavior of the Poisson's ratio for the armchair and the zigzag graphene sheets under large strain qualitatively agrees well with the trend observed in the existing DFT and molecular dynamics studies conducted in Refs. [39, 50]. We can observe that the Poisson's ratio is almost constant under small strains up to 1.2%, but decays differently under large strains. Our results also indicate that the graphene sheet behaves in an isotropic manner under small strains (< 1.8%) because the 2D hexagonal symmetry of pure C lattice ensures its isotropic elastic properties.

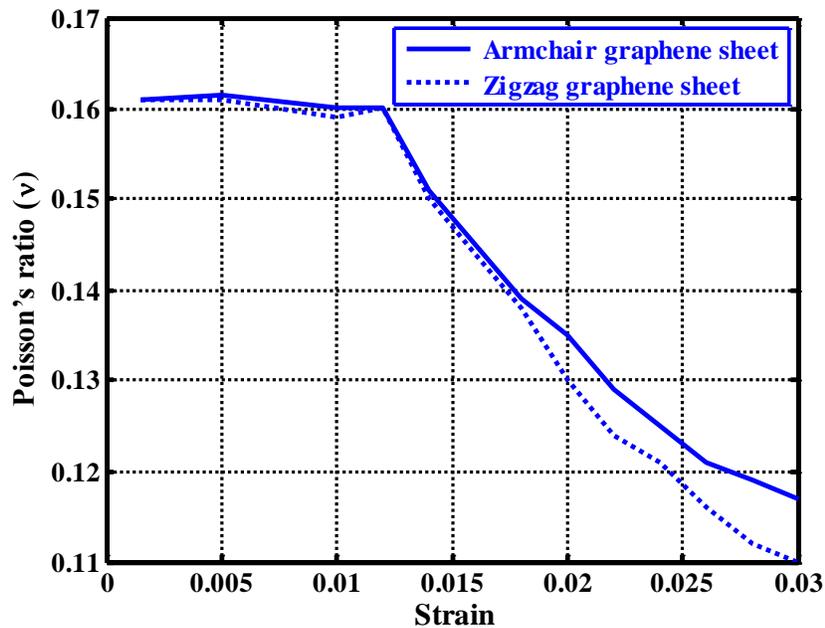

**Fig. 6.** Poisson's ratio of armchair and zigzag graphene sheets.

Figure 7 demonstrates the variations of $E_{gap}$ for armchair and zigzag graphene sheets obtained from DFT and TB models for various strains up to 3%. In the TB modeling, we adopted $t_0 = 2.67$ eV and $C_t = 1.29$ [40,41] and used the Poisson's ratio determined from DFT calculations (see Fig. 6). Figure 7 clearly shows that asymmetrical strain distributions in graphene sheets result in the opening of the band gaps at the Fermi level and these band gaps increase as the strain increases. For the first case (see Fig. 2a), we found that the band gap of graphene does not open irrespective of the magnitude of symmetrical strain distribution and it remains always a semiconductor. This finding is consistent with a previously reported results on graphene [38]. It may also be observed from Fig. 7 that the values of $E_{gap}$ for armchair and zigzag graphene sheets are almost the same and that negligible differences attributed to different Poisson's ratio (when graphene is strained in one direction, it will shrink in the perpendicular direction) leads to a further shift the Fermi points. Figure 7 clearly shows good agreement between the TB modeling results and DFT calculations, validating our simulations and enabling us to determine uniaxial strain gradient induced polarization in graphene sheets with non-centrosymmetric trapezoidal shaped pore.

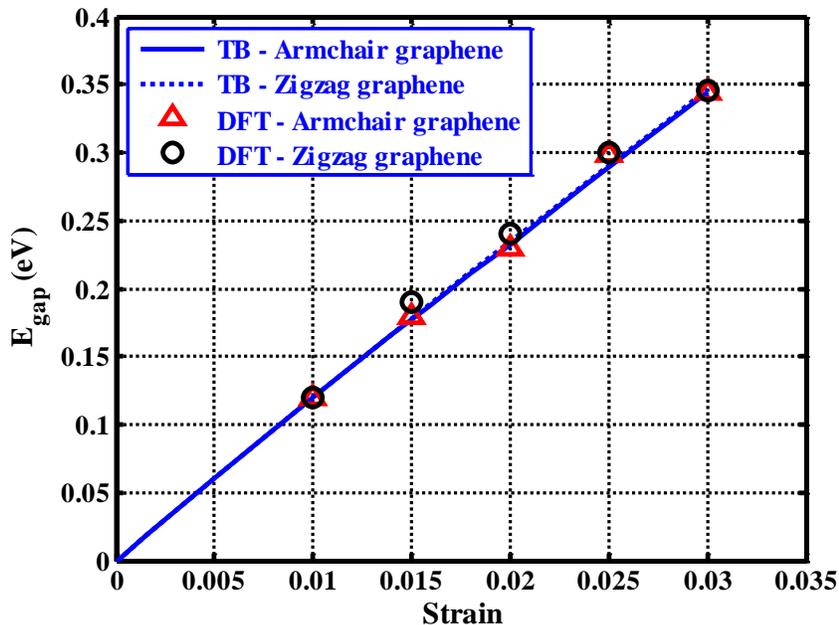

**Fig. 7.** Variations of $E_{gap}$ for armchair and zigzag graphene sheets subjected to uniaxial strain.

To further validate our quantum mechanics simulations, the experimentally determined dipole moments of carbonyl sulfide and aniline were considered. DFT calculations for carbonyl sulfide and aniline were performed using a generalized gradient approximation of Perdew-Burke-Ernzerhof [51]. Molecular systems of carbonyl sulfide and aniline were modeled by the ultrasoft pseudopotential method using the PWscf code from the Quantum ESPRESSO software package [45]. Orbitals were expanded in a plane-wave basis set up to a kinetic energy cutoff of 400 eV; 3000 eV for the charge density cutoff. Initial configurations of carbonyl sulfide and aniline structures were prepared and relaxed. Then, BZ integrations were performed with Gaussin-smearing [52] and Monkhorst-Pack k-point technique using a smearing parameter of 0.04 eV and a $4 \times 4 \times 4$ mesh for both structures. Table 1 summarises the outcome of this comparison. The two sets of results predicted are in good agreement and validate our simulations.

**Table 1.** Comparisons of molecular electric dipole moments

| Structure | Experimental | Present calculations |
|---|---|---|
| Carbonyl Sulfide | $0.71521 \pm 0.00020$ D* [53] | 0.715 |
| Aniline | $1.129 \pm 0.005$ D ($0^+$ state) [54] | 1.125 |
|  | $1.018 \pm 0.005$ D ($0^-$ state) [54] | 1.014 |

* 1D = $3.33564 \times 10^{-30}$ C m

## *4.1. Uniaxial strain induced piezoelectric effect*

In this section, we report results concerned with uniaxial strain induced polarization in armchair and zigzag graphene sheets containing non-centrosymmetric trapezoidal and triangular pores. An additional aspect of pore edge states by hydrogen termination is related to the current study because such states may alter the flexoelectric phenomena in graphene sheets. Figures 4 and 8 demonstrate passivated armchair graphene sheets with edge and corner (red circled) states.

Therefore, we focus on examining the effect of functionalization on strain induced polarization, considering not only the edge but also the corner states. Note that no additional force acts on the carbon atoms due to H-H repulsion along the zigzag direction. However, inhomogeneous force acts on the carbon atoms along the armchair direction, which causes lattice relaxation and, hence, interatomic bonding energy weakening [55, 56]. It may be observed from Figs. 4 and 8 that the passivated armchair graphene sheets under uniaxial strain always have trapezoidal pore with a modified zigzag edge. However, repulsion induced by hydrogen termination at the corner of a pore (red circled), which we call *corner strain effect*, may result. Accordingly, two cases were considered: two armchair graphene sheets with 4.5% and 20% atom vacancy concentrations (ρ) forming non-centrosymmetric trapezoidal pores with and without corner strain effect. The predicted strain gradient induced polarization in these graphene sheets as a function of strain is shown in Fig. 9. This figure clearly depicts that the graphene sheet with non-centrosymmetric trapezoidal shaped pore subjected to axial stress exhibits the required polarization, since the net average polarization is nonzero. The strain gradient induced polarization increases with the trapezoidal pore size. This is attributed to the strong dielectric behavior of a graphene sheet with larger pores. It may also be observed from these figures that the corner strain effect marginally affects the strain gradient induced polarization in case of armchair graphene sheet containing small pores. As expected, the reverse is true in case of armchair graphene sheet containing big pores. Figures 9(a) and 9(b) depict linear piezoelectric responses. In the case where corner strain effects are not present, the respective axial piezoelectric coefficients were determined from these linear least square fits to be $0.027 \text{ C/m}^2$ and $0.12 \text{ C/m}^2$; while in the case involving corner strain effects, the corresponding values are determined to be $0.023 \text{ C/m}^2$ and $0.115 \text{ C/m}^2$. The full piezoelectric tensor for graphene sheet can be determined on the basis of symmetry of hexagonal $\bar{6}m2$ class using the relations: $e_{12} = -e_{11}$ and $e_{26} = -e_{11}$.

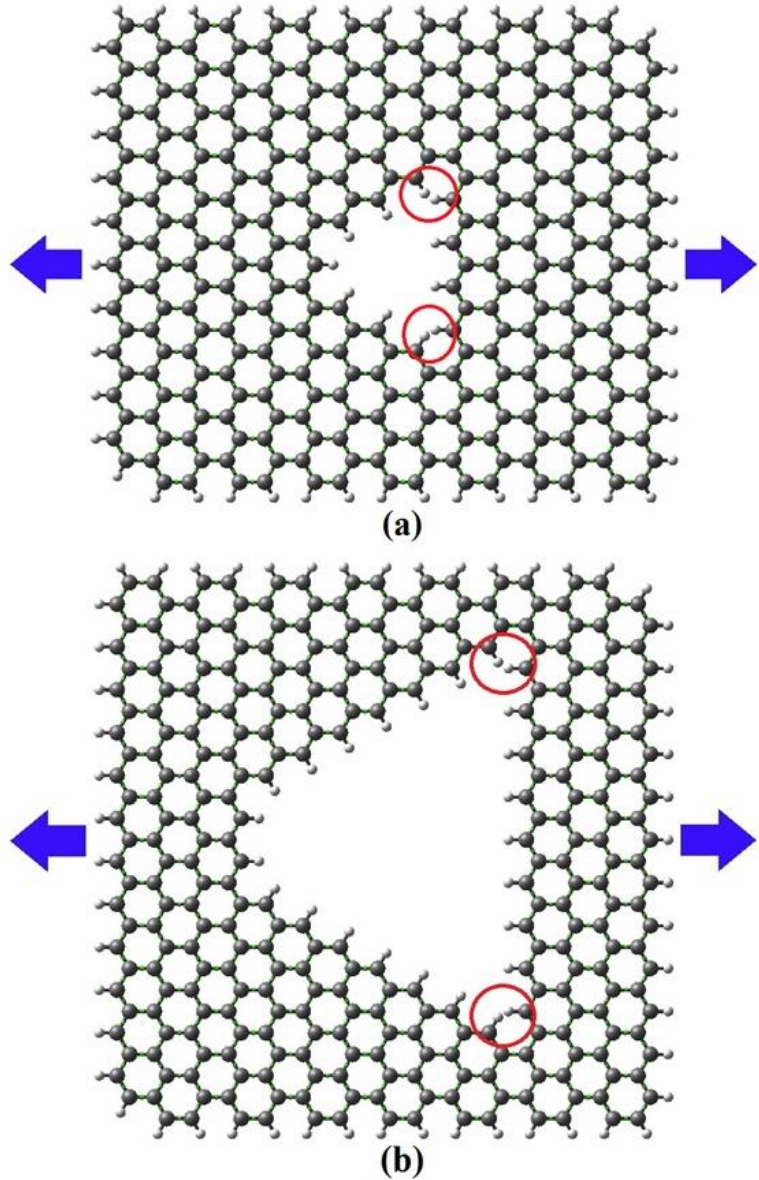

**Fig. 8.** Passivated armchair graphene sheet with trapezoidal pore subjected to an axial stress: (a) atom vacancy concentration of 4.5%, and (b) atom vacancy concentration of 20%.

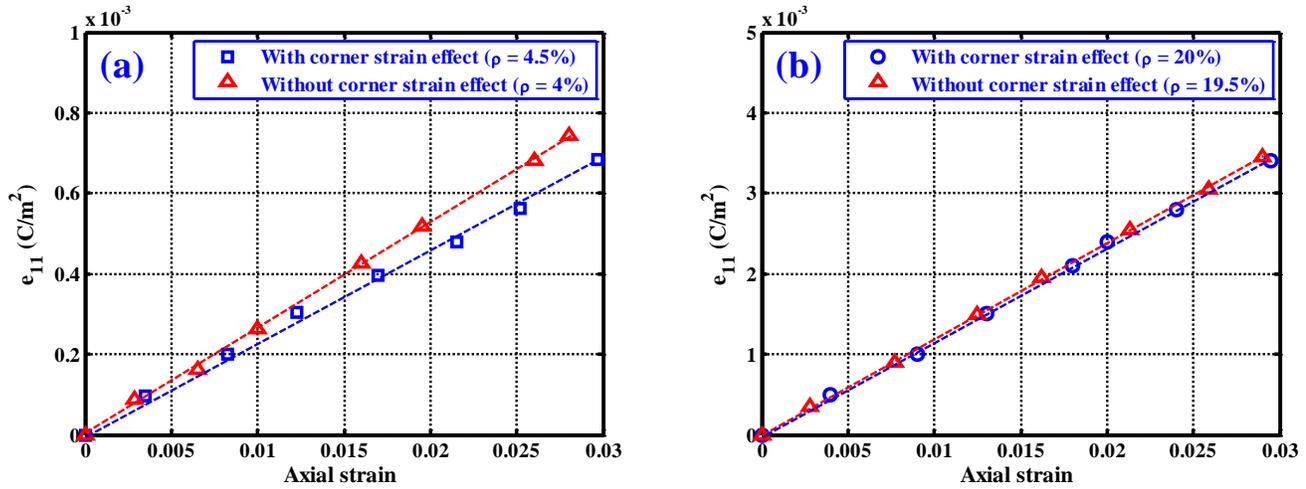

**Fig. 9.** Strain induced average polarization in armchair graphene sheets containing non-centrosymmetric trapezoidal pore with and without corner strain effects.

As mentioned earlier, certain symmetry rules for the pores should be followed to induce polarization in the graphene sheet. Therefore, we endeavoured to establish whether other pore geometries induce strain polarization in a graphene sheet. For such an investigation, we considered a triangular pore in a graphene sheet, as shown in Fig. 10. This figure clearly indicates that an armchair graphene sheet with a triangular pore is more sensitive to corner strain effects. Note that the armchair graphene sheet containing a triangular pore without corner strain effects represent the earlier discussed case of armchair graphene sheet containing trapezoidal pore (see Fig. 4a). We compared these two cases of armchair graphene sheets with 4.8% and 4% atom vacancy concentrations forming non-centrosymmetric triangular pores with and without corner strain effect. The predicted strain gradient induced polarization in these graphene sheets as a function of strain is shown in Fig. 11. We can clearly observe from this figure that the induced polarization increases with the applied strain. It may also be observed that corner strain effects significantly affect the strain gradient induced polarization even if the vacancy concentration is maximum in the case of a triangular pore. Note that corner strain effects become less pronounced in case of a larger pore size, as we have seen in the earlier case of a trapezoidal pore. The respective axial piezoelectric coefficients were determined from linear least square fits to be 0.0216 $C/m^2$ and 0.27 $C/m^2$ with and without corner strain effects. It is important to note from Figs. 9 and 11 that the corner strain effects can be systematically controlled in case of small pores in armchair sheets by adopting various symmetry rules and can be neglected in case of larger pores.

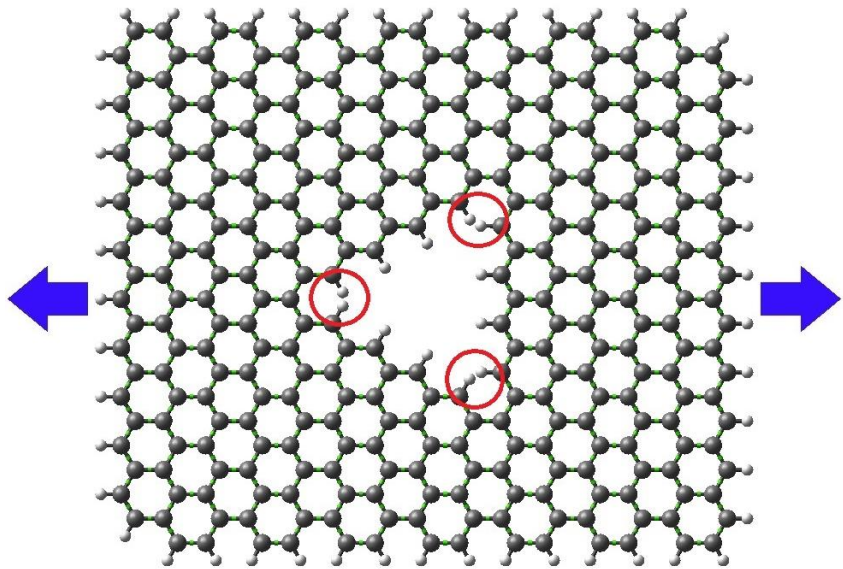

**Fig. 10.** Passivated armchair graphene sheet with triangular pore subjected to an axial stress (4.8% atom vacancy concentration).

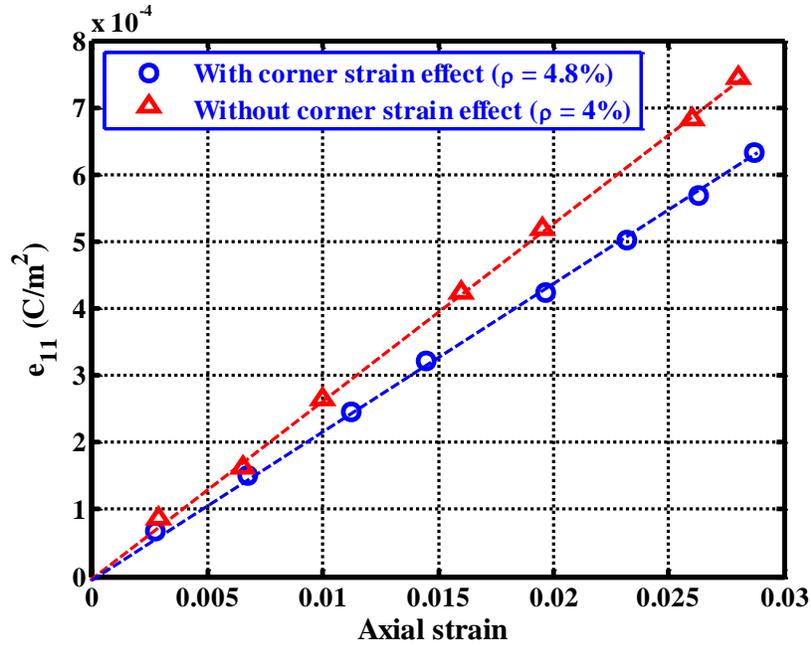

**Fig. 11.** Strain induced average polarization in armchair graphene sheets containing non-centrosymmetric triangular pore with and without corner strain effect.

In the previous sets of results, the strain induced polarization has been investigated by applying the tensile load along the armchair direction of graphene sheets. However, the application of the tensile load along the zigzag direction of graphene sheet may affect the strain induced polarization in it because of the ever-present existence of H-H repulsion along the armchair edges of a pore (elliptically marked region in blue color) and corner strain effects (red circled), as shown in Fig. 12. Figure 13 shows the predicted strain gradient induced polarization in zigzag graphene sheets as a function of the applied strain. A similar trend of results is obtained for the values of $e_{11}$ as has already been observed in case of armchair graphene sheets but in case of zigzag graphene sheet, the strain gradient induced polarization is much less than that found in case of former with less vacancy concentration. For instance, the values of axial piezoelectric coefficients are 0.031 C/m² and 0.034 C/m² for the zigzag graphene sheets with 7.5% and 8.2% atom vacancy concentrations, respectively.

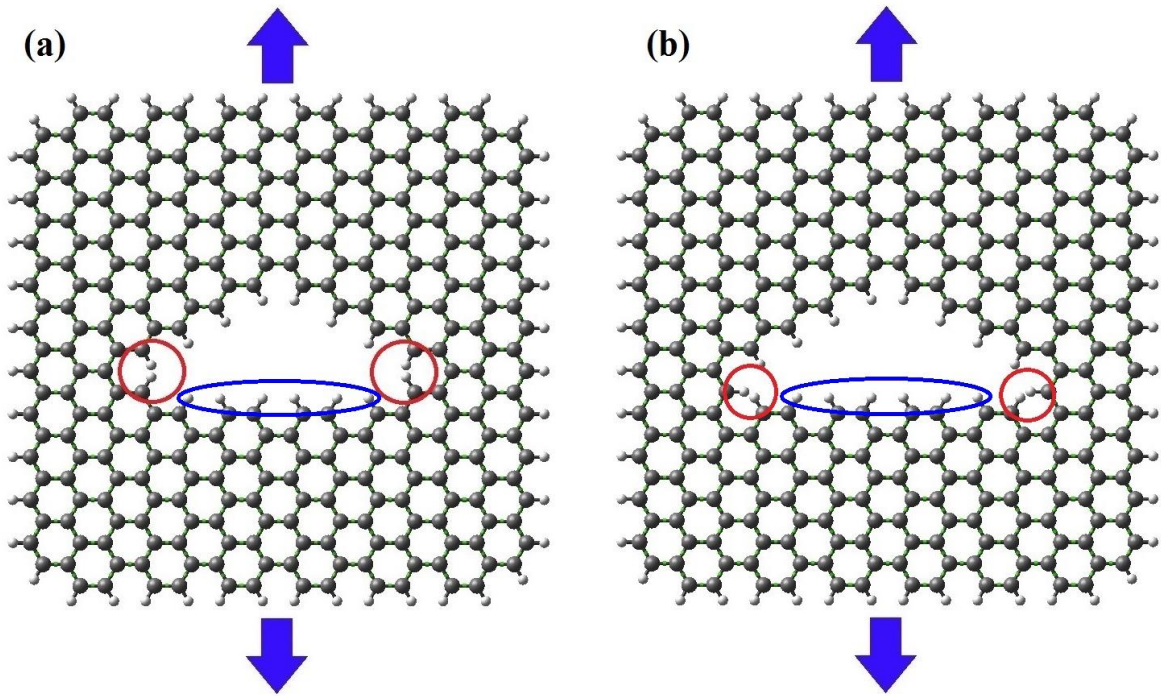

**Fig. 12.** Passivated zigzag graphene sheet with trapezoidal pore subjected to an axial stress: (a) atom vacancy concentration of 7.5%, and (b) atom vacancy concentration of 8.2%.

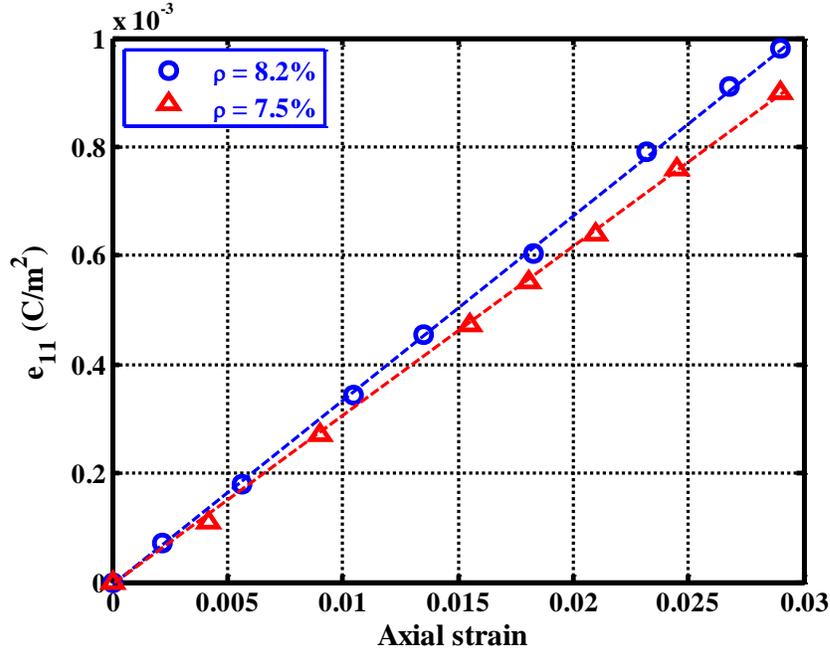

**Fig. 13.** Strain induced average polarization in zigzag graphene sheets containing non-centrosymmetric trapezoidal pore with the ever-present corner and edge strain effects.

## *4.2. Curvature induced piezoelectric effect*

Next, we investigated the curvature-induced polarization into the rolled graphene sheet by determining the magnitude of the induced normal atomic dipole at each atomic site. Figure 14 shows the dipole moment per atom for four different radii of curvature. As expected, the magnitude of the dipole moment per atom decreases as the radius of curvature increases. Note that the graphene layer can sustain large deformations as witnessed by their ability to form near perfect carbon nanotube. Some studies suggest that curvature induced charge redistribution is presents in carbon nanotube which is non-existent in flat graphene sheet [32]. A build-in electrostatic field perpendicular to CNTs surface exists due to its curved geometry. The surface electric field is believed to cause antilocalization, whose signature is seen in the reported positive magnetoresistance measurements in CNTs under low magnetic field. It is worth noting that the piezoelectric coefficients are given per unit volume. Therefore, a quantitative determination of the piezoelectric properties of the 2D graphene sheet to the 3D equivalent graphene layer requires specification of the interlayer spacing and packing, representing the electron distribution around it. We assumed 2D graphene sheet as an equivalent 3D graphene layer using an assumed interlayer separation of 3.4 Å as its thickness [57-60]. Once the dipole moment per atom for different radii of curvature of graphene is obtained, we can determine the dipole density (i.e., average curvature-induced polarization) using the following relation:

$$e_{33} = \frac{\text{dipole moments of n number of atoms}}{\text{volume occupied by n number of atoms}} \qquad (19)$$

Note that C-C bond length was taken as 1.41 Å while computing the volume of an equivalent 3D graphene layer. Figure 15 demonstrates the normal polarization in a graphene sheet for four different radii of curvature as a function of number of atoms (n) in it. We considered a square shaped graphene made of n atoms to investigate the effect of its different radii of curvature. In general, the obtained values of normal piezoelectric constants are greater than Gallium nitride (0.73 $C/m^2$) [61], α-Quartz (0.18 $C/m^2$) [62], and piezoelectric polymer poly(vinylidene fluoride) (0.168 $C/m^2$) [63], which are all commonly used as piezoelectric materials. Alternatively, the piezoelectric constants computed from different dimensional system can be expressed as a total dipole per stoichiometric unit. The full piezoelectric tensor for graphene sheet can be determined on the basis of symmetry of hexagonal $\bar{6}m2$ class using the relations: $e_{31} = -e_{33}$ and $e_{15} = -e_{33}$.

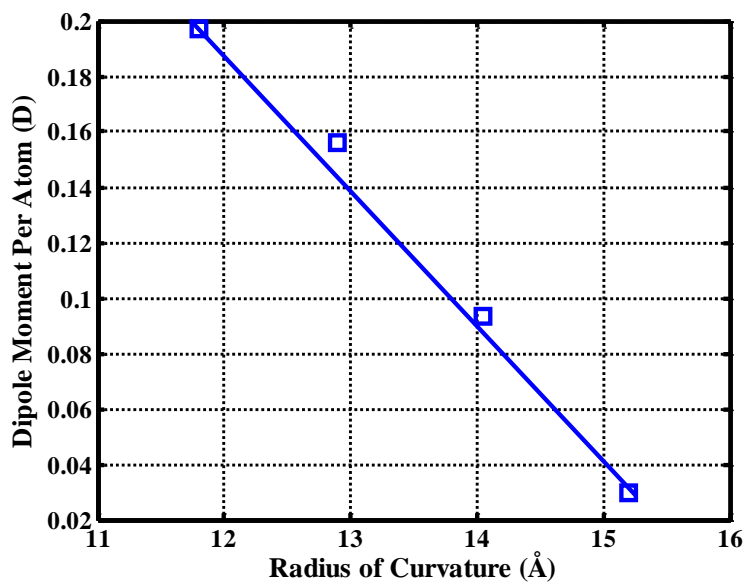

**Fig. 14.** Evolution of dipole moment per atom as a function of radius of curvature (R) of graphene.

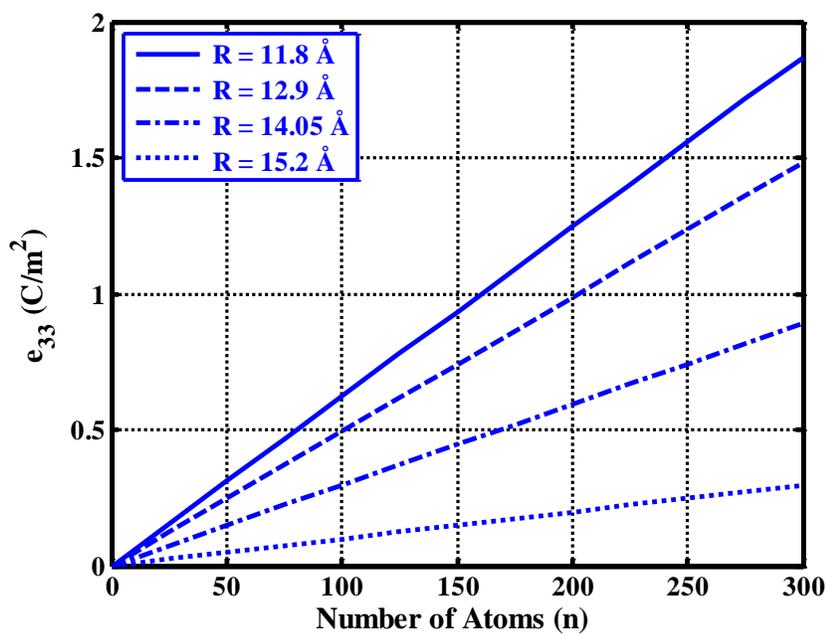

**Fig. 15.** Curvature induced normal piezoelectric coefficient ($e_{33}$) as a function of number of atoms (n) in graphene.

## 5. Conclusions

In this study, we report the induced piezoelectricity in graphene sheets through interplay between different non-centrosymmetric pores with edge and corner states, curvature, and flexoelectricity phenomena in the presence of strain gradients using quantum mechanics calculations. We have also calculated the electronic structure of armchair and zigzag graphene sheets under planner asymmetrical strain distributions using quantum mechanics calculations and tight-binding approach, which result in the opening of band gaps at the Fermi level. Our results suggest that a graphene sheet with non-centrosymmetric trapezoidal-shaped pores and bent graphene show strong electromechanical coupling when subjected to large strain gradients. We also studied the effects of functionalization, considering not only the edge state but also the corner state. We found that zigzag graphene sheets provide weaker flexoelectric effects in comparison with armchair graphene sheets for the same vacancy concentration. This is due to the ever-present existence of H-H repulsion along the armchair edge of a pore and corner strain effects. In the case of armchair graphene sheets, the corner strain effects can be systematically controlled, or even eliminated by adopting various symmetry rules and can be neglected in cases involving larger pores. The current study offers different pathways to alter the axial and normal electromechanical coupling in graphene sheets through changing the size of the non-centrosymmetric pores and the radius of curvature of these sheets. In particular, the latter case provides a strong electromechanical coupling without creating pores in the graphene sheet and possibly sacrificing its elastic properties and mechanical strength.

**Acknowledgements:**

This work was supported by the Natural Sciences and Engineering Research Council of Canada

(NSERC) and a Banting Postdoctoral Fellowship awarded to the first author. The authors wish to acknowledge the anonymous reviewers for their constructive remarks and suggestions.